\documentclass[runningheads]{llncs}

\usepackage[T1]{fontenc}
\usepackage[utf8]{inputenc}
\usepackage{acro}
\usepackage{adjustbox}
\usepackage{booktabs}
\usepackage{comment}
\usepackage{todonotes}
\usepackage{colortbl}

\newcolumntype{a}{>{\columncolor{gray!10!white}}c}
\newcolumntype{x}{>{\columncolor{green!10!white}}c}
\newcolumntype{y}{>{\columncolor{blue!10!white}}c}
\newcolumntype{z}{>{\columncolor{yellow!10!white}}c}
\newcolumntype{v}{>{\columncolor{red!10!white}}c}
\definecolor{OliveGreen}{rgb}{0,0.6,0}
\definecolor{ForestGreen}{RGB}{34,139,34}
\definecolor{myblue}{RGB}{37,165,203}
\definecolor{FAUblue}{rgb}{0.000, 0.2196, 0.3961}
\definecolor{myred}{RGB}{175,32,67}

\colorlet{backgroundcol}{cyan!10!white}
\newcommand{\highlight}[1]{%
	\par\noindent
	\fcolorbox{black}{backgroundcol}{%
		\parbox{\dimexpr\linewidth-2\fboxsep\relax}{%
			#1
		}%
}}

\usepackage{graphicx}
\usepackage{listings}
\usepackage{hyperref} 
\begin{document}
%

\title{Leveraging HPC Profiling \& Tracing Tools to Understand the Performance of Particle-in-Cell Monte Carlo Simulations}
\titlerunning{Understanding the Performance of Particle-in-Cell Monte Carlo Simulations}
%

\author{Jeremy J. Williams\inst{1}\and
David Tskhakaya\inst{2}\and
Stefan Costea\inst{3}\and
Ivy B. Peng\inst{1}\and
Marta Garcia-Gasulla\inst{4}\and
Stefano Markidis\inst{1} }
\authorrunning{Jeremy J. Williams et al.}

%
\institute{KTH Royal Institute of Technology, Stockholm, Sweden \and Institute of Plasma Physics of the CAS, Prague, Czech Republic \and
LeCAD, University of Ljubljana, Ljubljana, Slovenia \and
Barcelona Supercomputing Center, Barcelona, Spain}
\maketitle              
%

\begin{abstract}
Large-scale plasma simulations are critical for designing and developing next-generation fusion energy devices and modeling industrial plasmas. BIT1 is a massively parallel Particle-in-Cell code designed for specifically studying plasma material interaction in fusion devices. Its most salient characteristic is the inclusion of collision Monte Carlo models for different plasma species. In this work, we characterize single node, multiple nodes, and I/O performances of the BIT1 code in two realistic cases by using several HPC profilers, such as perf, IPM, Extrae/Paraver, and Darshan tools. We find that the BIT1 sorting function on-node performance is the main performance bottleneck. Strong scaling tests show a parallel performance of 77\% and 96\% on 2,560 MPI ranks for the two test cases. We demonstrate that communication, load imbalance and self-synchronization are important factors impacting the performance of the BIT1 on large-scale runs. 
\end{abstract}


\section{Introduction}
Plasma simulations are a key asset and tool for improving current and next-generation plasma-based technologies, such as fusion devices, and industrial applications, such as plasma lithography for chip production, to mention a few examples. The Particle-in-Cell (PIC) methods are the most widely used numerical technique to plasmas from first principles.  BIT1 is a massively parallel PIC code for studying complex plasma and their interaction with materials~\cite{tskhakaya2010pic}. Its main feature is modelling plasma bounded between two conducting walls with an applied external circuit and inclusion of collisions.
This work's primary focus is investigating, characterizing and understanding the performance of BIT1. To achieve this, we use several profiling and tracing tools and analyze the results. The main contributions of this work are the following:

\begin{itemize}
    \item We identify the most computationally intensive parts of the BIT1 code, which are amenable for performance optimization, and analyse the performance of running BIT1 on a single node.
    \item We apply profiling and tracing techniques to evaluate the MPI communication cost, load balancing, parallel efficiency and I/O performance in strong scaling tests.
\end{itemize}


\section{PIC/Monte Carlo BIT1 Simulation Code}
In its first inception, the BIT1 code is a 1D3V PIC code: this means that simulations are performed in one-dimensional space using three dimensions for the particle velocities. While particles are only allowed to move in one direction, particles have three components for the velocities. 

\begin{figure}
    \begin{center}
        \includegraphics[width=0.85\textwidth]{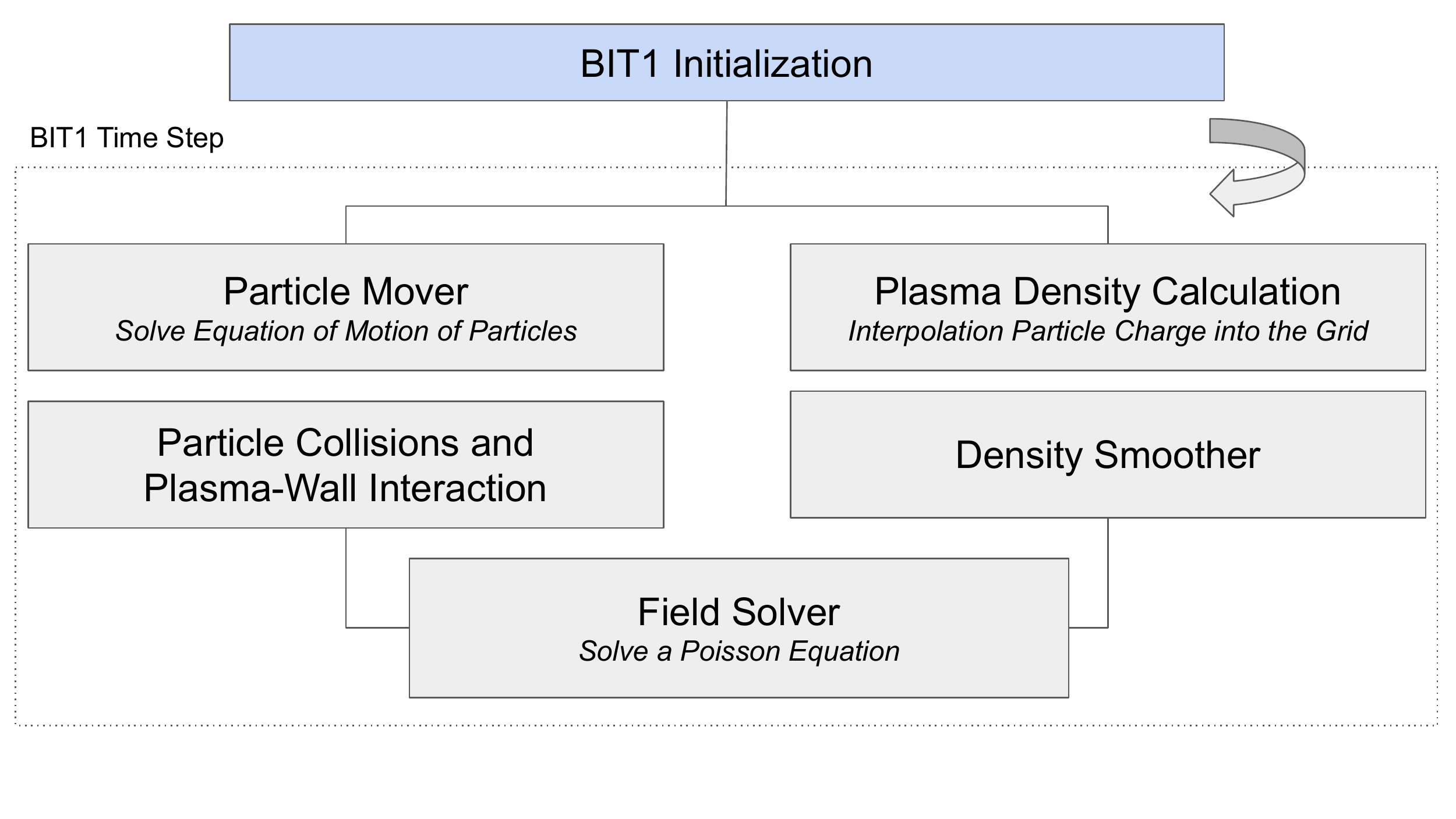}
        \caption{A diagram representing the algorithm used in BIT1.} \label{diagram}
     \end{center}
\end{figure}

The PIC method is one of the most widely used simulation techniques for plasma simulation. At its heart, it consists of calculating the particle trajectories for million particles in a field consistent with the density and current distributions, e.g., that satisfy the Maxwell's equations. BIT1 uses an explicit formulation of the PIC method. 
Fig.~\ref{diagram} shows the basic algorithm of the BIT1 code. First, we initialize the simulation by setting up the computational grid and particle (of different species, such as electrons and ions) position and velocity. Then a computational cycle is repeated several times to update the field (the Electric field) and particle position and velocity. BIT1 includes sophisticated Monte Carlo techniques to mimic collisions and ionization processes. At this end, BIT1 can simulate considerably realistic configurations relevant to plasmas in the laboratories, including fusion machines. 


A few distinctive phases are carried out in each simulation time step (typically, there are a hundred thousand steps).

As shown in Fig.~\ref{diagram}, the computational cycle consists of five phases: (i) plasma density calculation using particle to grid interpolation (ii) a density smoother to remove high-frequency spurious frequencies iii) a field solver using a liner system (iv) calculation of particle collisions and interaction with the walls using a Montecarlo technique (v) a particle mover to advance in time the particle positions and velocities.
Together with these five phases, there is an I/O phase occurring only for specific time steps (for instance, only every 1,000 cycles), enabling diagnostics and providing checkpointing and restart capabilities.
 
BIT1 is a successor of the \texttt{XPDP1} code, developed by Verboncoeur's team at Berkeley~\cite{verboncoeur1993simultaneous}. It comprises approximately 31,600 lines and is entirely written in the C programming language. Currently, BIT1 is not relying on numerical libraries but natively implements solvers for the Poisson equation, particle mover, and smoother.

The main BIT1 feature is the data layout for storing the particle information, as explained in Ref.~\cite{tskhakaya2007optimization}. While in the typical, simple PIC formulation, the data of \texttt{i}-th particles are stored as an array \texttt{A[s][i]}, where \texttt{A} is the particle property (like coordinate, or velocity components) and \texttt{s} is the particle species, BIT1 associates the particles with the cells they are located in and stores the particle information as \texttt{A[s][k][i]}, where \texttt{k} is the cell number where the particle is at a given time. As we know \texttt{k}, then \texttt{A[s][k][i]} can represent the relative position inside the \texttt{k} cell (and we can also use single-precision floating points). This particle data layout has the advantage that particles neighbouring in space are also neighbours in memory, increasing the possibility of having cache hits. Another advantage of this approach is that it allows for easier development of Montecarlo collision models.

However, one of the disadvantages of this approach is the fact that the particles needs to be sorted: for each particle, we need to \emph{i)} check the new position of the particle, and \emph{ii)} if a given particle has left its cell, update its corresponding address accordingly. The function responsible for this in BIT1 is called \texttt{arrj}.

BIT1 uses a domain decomposition for parallelization and MPI for parallel communication. The one-dimensional grid is divided across different MPI processes, and MPI point-to-point communication is used for halo exchange for the smoother, Poisson solver and particles (exiting the domain). BIT1 uses point-to-point non-blocking communication.


\section{Methodology \& Experimental Setup}
This work aims to understand the BIT1 performance bottleneck and potential improvements. We use several performance analysis tools, such as profilers and tracers, to achieve this. In particular, we use the following tools:
\begin{itemize}
    \item \texttt{gprof} is an open-source profiling tool that gathers information on the execution time and reports the relevant functions used most often by the processor. Since each MPI process produces a \texttt{gprof} output, the different profiling information is reduced to one containing all statistics. 
    \item \texttt{perf} is a low-level profiler used to gather hardware performance counters. In particular, we use \texttt{perf} to characterize the cache and memory system,  performance.
    \item \texttt{IPM} or Integrated Performance Monitoring is a profiling performance monitoring tool that captures the computation and communication of a parallel program. Specifically, \texttt{IPM} reports on MPI calls and buffer sizes~\cite{fuerlinger2010effective}. We use \texttt{IPM} to understand the parallel and MPI performance and evaluate the impact of workload imbalance.
    
    \item \texttt{Extrae} \& \texttt{Paraver} are parallel performance tracing and profiler tools, developed by the Barcelona Supercomputing Center (BSC)~\cite{servat2013framework}. Specifically, \texttt{Extrae} is used to instrument the code and Paraver to post-process the \texttt{Extrae} output and visualize it. 
    \item \texttt{Darshan} is a performance monitoring tool, specifically designed for analysing serial and parallel I/O workloads~\cite{snyder2016modular}. We use \texttt{Darshan} to evaluate the I/O performance of BIT1 in terms of write bandwidth.
\end{itemize}

%

\subsection{BIT1 Test Cases}
We monitor and analyse BIT1 performance on two test cases that differ in problem size and BIT1 functionalities. We consider two cases: \emph{i)} a relatively straightforward run simulating neutral particle ionization due to interaction with electrons, and \emph{ii)} formation of high-density sheath in front of so-called \emph{divertor} plates in future magnetic confinement fusion devices, such as the ITER and DEMO fusion energy devices. More precisely, the two cases are as followed:
\begin{itemize}
    \item \textbf{Neutral Particle Ionization Simulation}. In this test case, we consider unbounded unmagnetized plasma consisting of electrons, $D^+$ ions and $D$ neutrals. Due to ionization, neutral concentration decreases with time according to $\partial n / \partial t = n n_e R$, where $n$, $n_e$ and $R$ are neutral particles, plasma densities and ionization rate coefficient, respectively. We use a one-dimensional geometry with 100K cells, three plasma species ($e$ electrons, $D^+$ ions and $D$ neutrals), and 10M particles per cell per species. The total number of particles in the system is 30M. Unless differently specified, we simulate 200K time steps. An important point of this test is that it does not use the Field solver and smoother phases (shown in the diagram of Fig.~\ref{diagram}). The baseline simulation for this test case uses one node of the Dardel supercomputer with a total of 128 MPI processes.

    \item \textbf{High-Density Sheath Simulation.} We consider a double bounded magnetized plasma layer between two walls. Initially the system is filled by a uniform plasma consisting of electrons and $D^+$ Deuterium ions. During the simulation, plasma is absorbed at the walls initiating recycling of $D$ neutrals and plasma sheath is forming. We use a one-dimensional geometry with three million cells, three plasma species ($e$ electrons, $D^+$ ions and $D$ neutrals), and 200M particles per cell per species. The total number of particles in the system is approximately 2.2B. Unless differently specified, we simulate 100K time steps. The baseline simulation for this test case uses five nodes of the Dardel supercomputer with a total of 640 MPI processes, since this simulation cannot fit in the memory of one node.
    
    
\end{itemize}








\subsection{Hardware and Software Environment}
We simulate and evaluate the performance of PIC/MC BIT1 on two systems: 

\begin{itemize}
\item \textbf{Greendog} is a workstation with an i7-7820X processor (8 cores), 32 GB DRAM, and one NVIDIA RTX2060 SUPER GPU.  The processor has a L1 cache 256 KiB size, L2 cache 8 MiB size and L3 cache (LLC) 11 MiB size.
\item \textbf{Dardel} is a HPE Cray EX supercomputer, with 554 compute nodes with 256GB DRAM and two AMD EPYC Zen2 2.25 GHz 64 core processors per node. The nodes are interlinked with a HPE Slingshot network using a Dragonfly topology and currently with a bandwidth of 100 GB/s. Each processor has a L1 cache 32 KiB size, L2 cache 512 KiB size and L3 cache (LLC) 16.38 MiB  size. The storage employs a Lustre file system. 
\end{itemize}

\section{Performance Results}
As a first step of this work, we analyse the impact of different compiler automatic optimization using flags -O0, -O2 and -O3 respectively.
\begin{figure}[h!]
    \begin{center}
        \includegraphics[width=\textwidth]{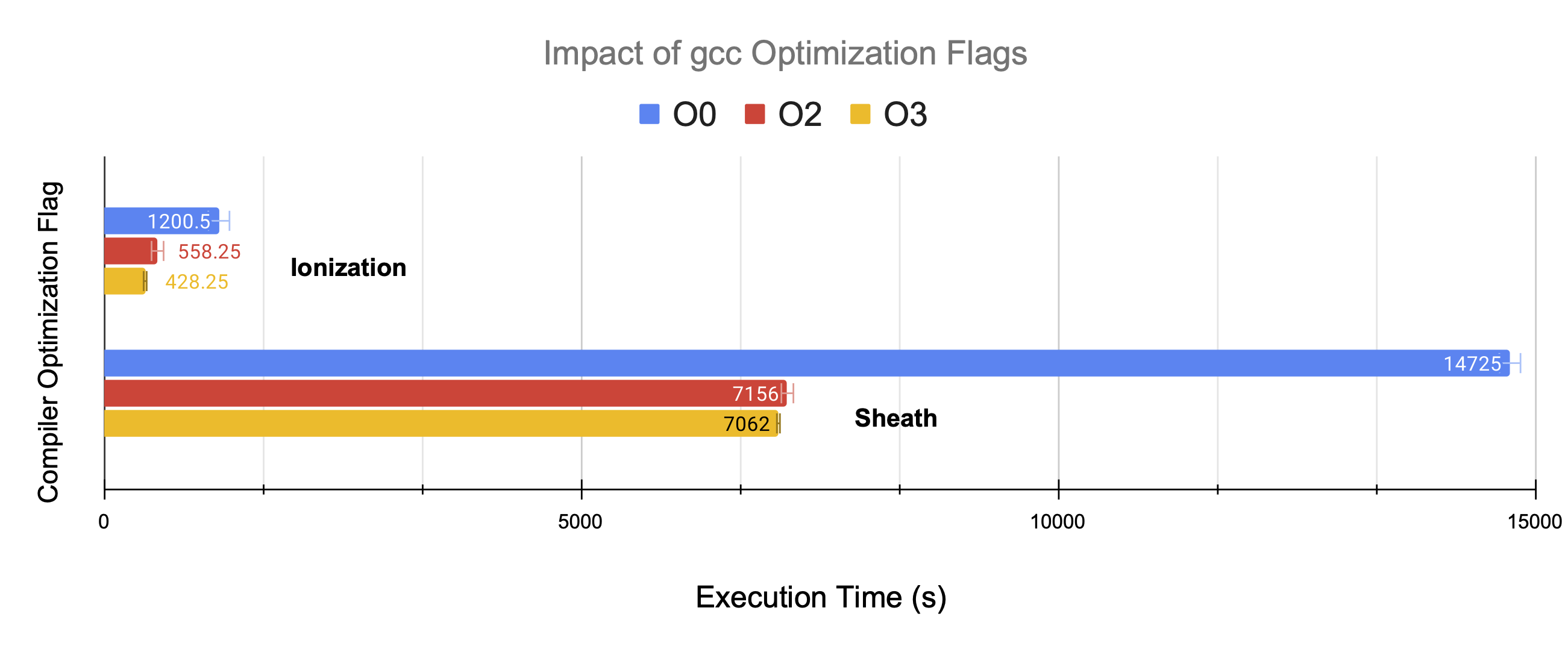}
        \caption{Impact of the \texttt{gcc} optimization flags for the Ionization and Sheath test cases.} \label{flags}
     \end{center}
\end{figure}
Fig.~\ref{flags} shows the execution time for BIT1 codes compiled with different compiler flags for the two test cases in the baseline configuration on the Dardel supercomputer. The -O2 and -O3 flag leads to an impressive performance improvement of more than 50\%. This is largely due to vectorization in the particle mover phase, where particle coordinates and velocities can be calculated in SIMD fashion thanks to the auto-vectorization. However, it's important to note that in some cases, -O3 optimizations may introduce subtle bugs or unexpected behavior due to the aggressive transformations applied to the code. Although such cases are rare, using the -O2 optimization is generally considered more stable. Therefore, for the remaining tests, we will use the -O2 optimization flag as -O3 optimization might be too aggressive.

As probably the most important part of the performance analysis, we identify which functions take most of the computing time. 
\begin{figure}[h!]
    \begin{center}
        \includegraphics[width=\textwidth]{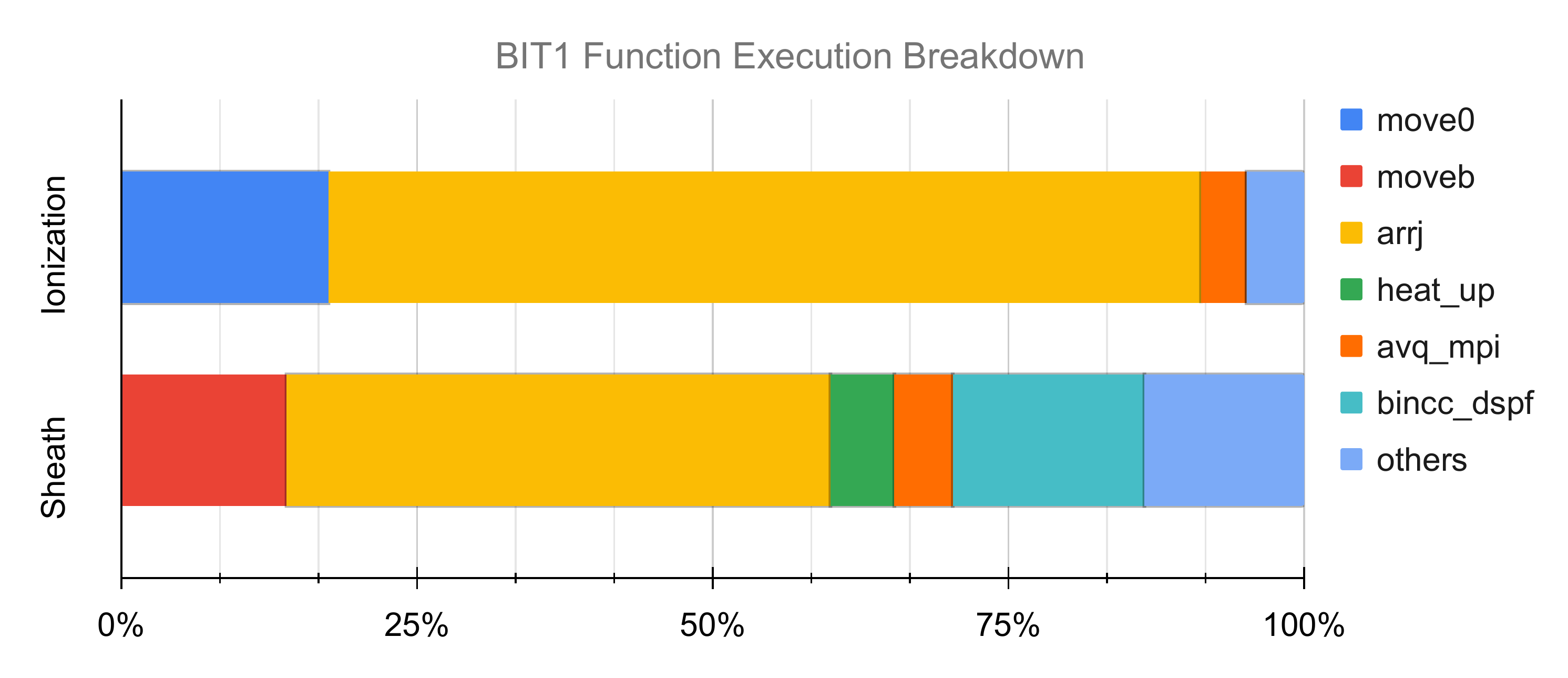}
        \caption{Percentage breakdown of the BIT1 functions where most of the execution time is spent for the Ionization and Sheath baseline cases. The \texttt{arrj} sorting function (in yellow colour) is the function that takes most of the time. The \texttt{gprof} tool have been used.} \label{breakdown}
     \end{center}
\end{figure}
Fig.~\ref{breakdown} shows the distribution of computational time in the two test cases using baseline configurations on Dardel. For the two test cases, the function taking most of the time is \texttt{arrj} (in the yellow color): \texttt{arrj} takes 75.5\% and 51.1\% of the total execution time, for the ionization and sheath test baseline cases on Dardel, respectively. 

\vspace{2mm} 
\highlight{\textbf{Observation I:} The BIT1 serial execution time is dominated by the sorting function \texttt{arrj} characterized by many memory transfers and little to no computing. BIT1 in the current implementation and data layout might not benefit from the usage of GPUs (given the low arithmetic intensity of the \texttt{arrj} function) but will improve when adding high-bandwidth memories.}
\vspace{3mm} 


To further investigate the BIT1 performance dependence on the memory system, we study the hardware profile counters using \texttt{perf} on the \texttt{Greendog} workstation (as we have root privileges on the machine). Table~\ref{table:1} shows the L1 and Last Level Cache (LLC) load misses percentage.  For this investigation, we use a baseline case with 10K time-steps for the Ionization case; we then reduce the problem size by 10\%, e.g. the number of cells and particles-per-cell parameter are decreased by a factor of ten, and then again by 20\% (the number of cells and particles-per-cell reduced by a factor of 20). The reason for that is, we want to analyze the impact of problem size on the usage of the cache system. In addition to our \texttt{perf} investigation, we perform additional profiling with a \texttt{cache-test}~\cite{cachetest}
 code that is known to have good cache performance and we take it as a reference point. 


\begin{table}[!ht]
    \centering
    \resizebox{12.25cm}{!} 
    { 
    \begin{tabular}{|c|c|c|c|}
    \hline
    \rowcolor{lightgray}
        Baseline Size & 10\% Reduction Size & 20\% Reduction Size & \texttt{cache-test} \\ \hline
    \rowcolor{lightgray}
        L1 Load Misses & L1 Load Misses & L1 Load Misses & L1 Load Misses \\ \hline
        \cellcolor{green!40} 3.43\% & \cellcolor{green!40} 2.51\% & \cellcolor{green!40} 2.17\% &  \cellcolor{green!40} 5.53\% \\ \hline
        \rowcolor{lightgray}
        LLC Load Misses & LLC Load Misses & LLC Load Misses & LLC Load Misses \\ \hline
        \cellcolor{red!40} 99.07\% & \cellcolor{yellow!40} 52.25\% & \cellcolor{yellow!40} 47.51\% & \cellcolor{green!40} 18.95\% \\ \hline
    \end{tabular}    
    }
    \caption{BIT1 L1 \& LLC load misses percentages.}\label{table:1}
\end{table}

While high L1 performance (low L1 miss rate) is observed for all our tests, the LLC performance for BIT1 is poor for the baseline case (99\% of the load are misses!). However, as soon as the problem becomes smaller and fits the LLC, we start observing an acceptable performance of the LLC.

\vspace{2mm} 

\highlight{\textbf{Observation II:} The BIT1 performance considerably depends on the problem size and effective LLC usage. In serial runs, the BIT1 is a highly memory-bound code.}
\subsection{Parallel Performance}
We analyze first the communication pattern using Extrae/Paraver, which allows us to trace the communication pattern precisely.  
\begin{figure}[h!]
    \begin{center}
        \includegraphics[width=0.8\textwidth]{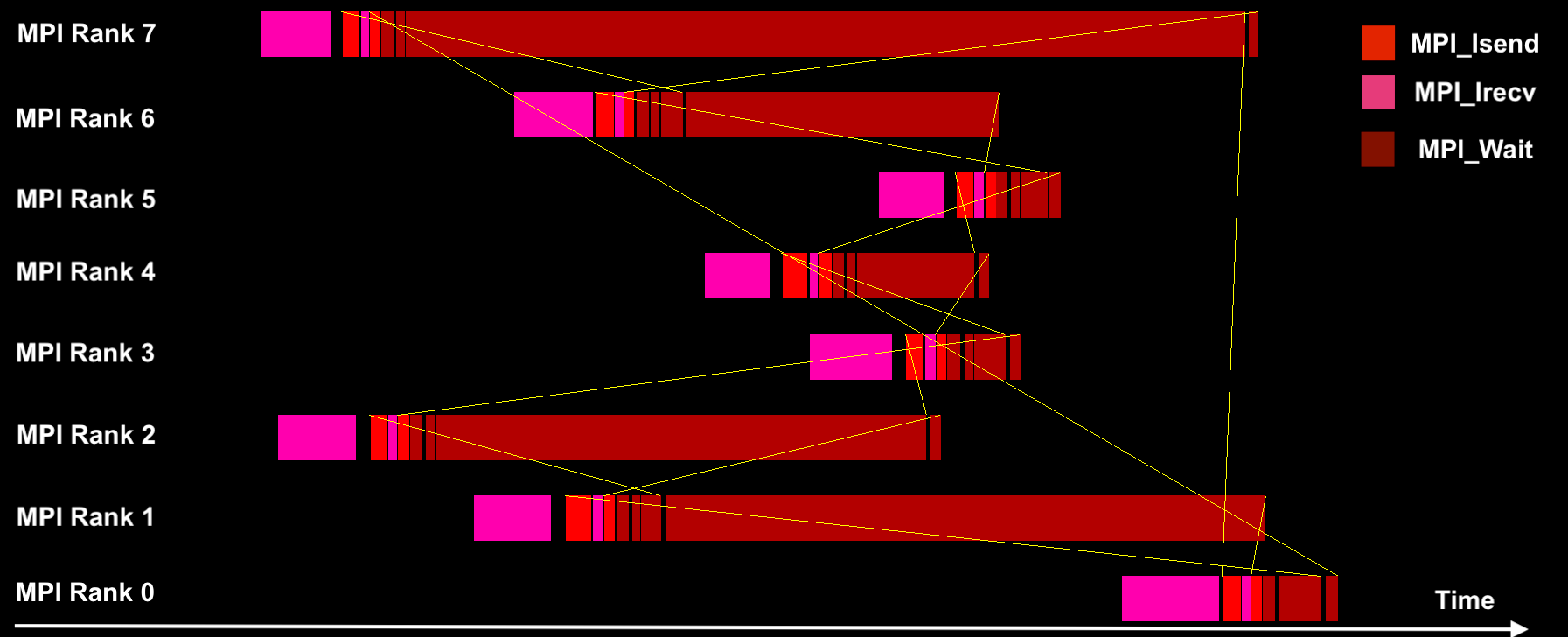}
        \caption{The MPI communication pattern is obtained, using \texttt{Extrae}/\texttt{Paraver}, with BIT1 using eight MPI processes on Dardel. The trace shows that MPI communication is non-blocking point-to-point and only involves neighboring processes. The MPI Rank \texttt{0} is the slowest. MPI ranks \texttt{1} and \texttt{7} wait for it, leading to a load imbalance.} 
        \label{runtimeParaver}
     \end{center}
\end{figure}
Fig.~\ref{runtimeParaver} shows a communication phase in BIT1. We note that the MPI communication is non-blocking point-to-point (\texttt{MPI\_Isend} / \texttt{MPI\_Irecv}) and only involves neighbouring processes (this is halo exchange in a one-dimensional domain-decomposition). The important point is that MPI rank 0 starts the communication late, e.g., it has more computation than other processes or is slower. Faster neighbour processes (MPI ranks \texttt{1} and \texttt{7}) must wait for MPI rank \texttt{0} to proceed. A simple tracing reveals that there is a potential imbalance situation.

\vspace{2mm} 

\highlight{\textbf{Observation III:} The trace of the parallel BIT1 communication shows the rise of workload imbalance, with the MPI rank \texttt{0} being the slowest and other neighbor MPI processes waiting for it as they are locally synchronized by the \texttt{MPI\_Wait} call.}

\vspace{2mm} 

\begin{figure} [h]
    \begin{center}
       \includegraphics[width=\textwidth]{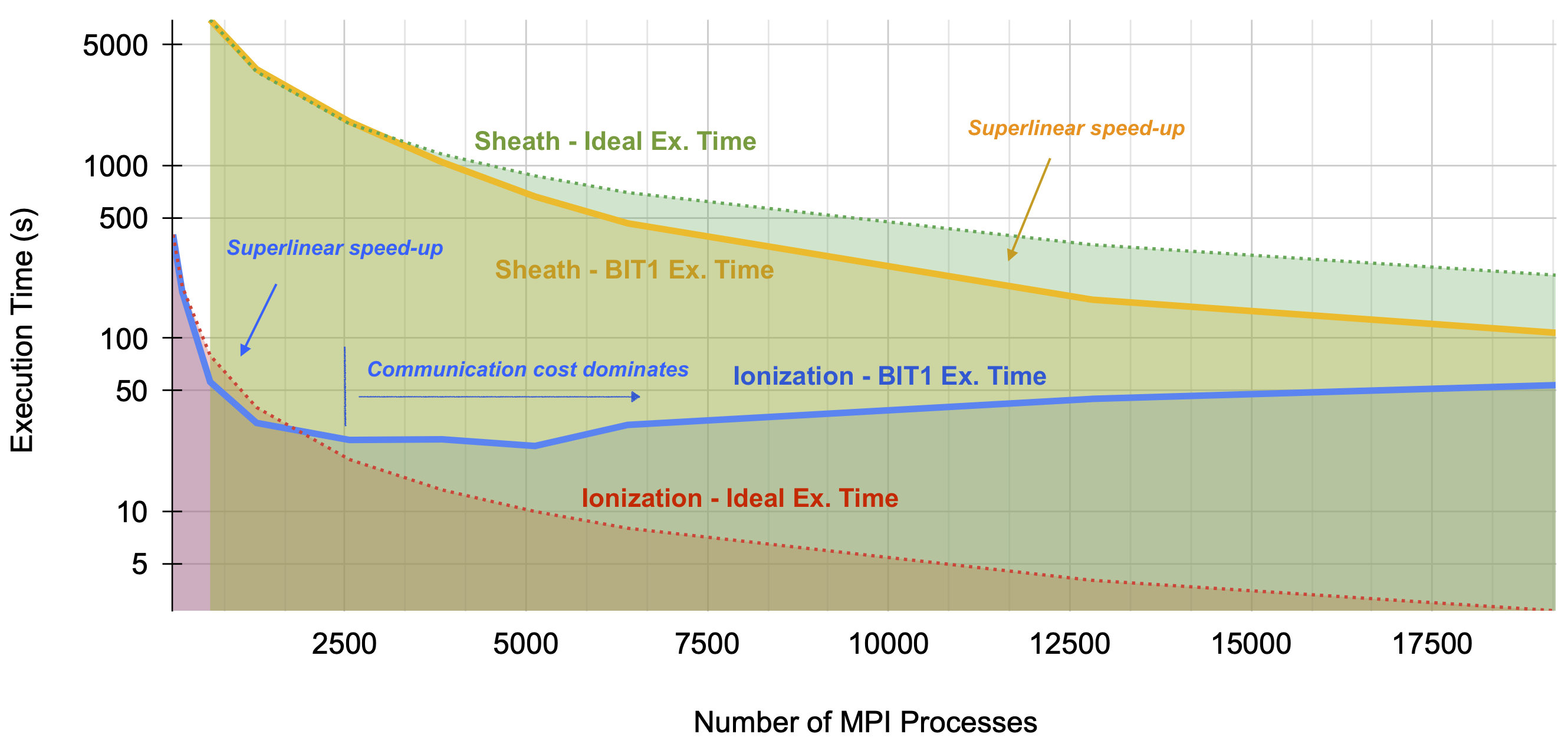}
        \caption{BIT1 strong scaling test execution times on Dardel supercomputer for the Ionization (blue line) and Sheath (yellow line). 
        } \label{scaling}
    \end{center}
\end{figure}

We then  perform a strong scaling study of the two simulation scenarios on Dardel. These are strong scaling tests as we fix the problem size and increase the number of cores in use. Fig.~\ref{scaling} shows the scaling of up to 19,200 cores. The Ionization simulation stops scaling at 2,560 cores, e.g., the problem size per MPI process is too small, and the communication cost takes large part of the simulation time: for instance, for the run on 19,200 cores, the communication takes 57\% of the total time with a large increase of the simulation!
On 2,560 MPI processes, the relative parallel efficiency is 77\% and 96\% for the Ionization and Sheath test cases, respectively. One important aspect of a strong scaling test is that BIT1 exhibits a superlinear speed-up~\cite{ristov2016superlinear} for both test cases. In particular, for the larger problem of the Sheath test case, when using more than 2,560 MPI processes, we always observe superlinear relative speed-up when comparing the speed-up to the five-node performance. The superlinear speed-up is because the problem size per MPI process decreases as the number of processes increases. As soon as the problem becomes small enough to fit into LLC, the performance vastly improves, leading to a superlinear speed-up.

\vspace{2mm} 

\highlight{\textbf{Observation IV:} BIT1 parallel performance shows a superlinear speed-up because increasing the number of MPI processes makes the problem size per process smaller and fits into LLC.}

\vspace{2mm} 

As the last part of the analysis of parallel communication, we study the load imbalance, as we suspected by analysing Fig.~\ref{runtimeParaver}, in the largest simulation we performed in the strong scaling test: 19,200 MPI processes on 150 nodes, using the Sheath test case. 

 \begin{figure} [!ht]
    \begin{center}
       \includegraphics[width=\textwidth]{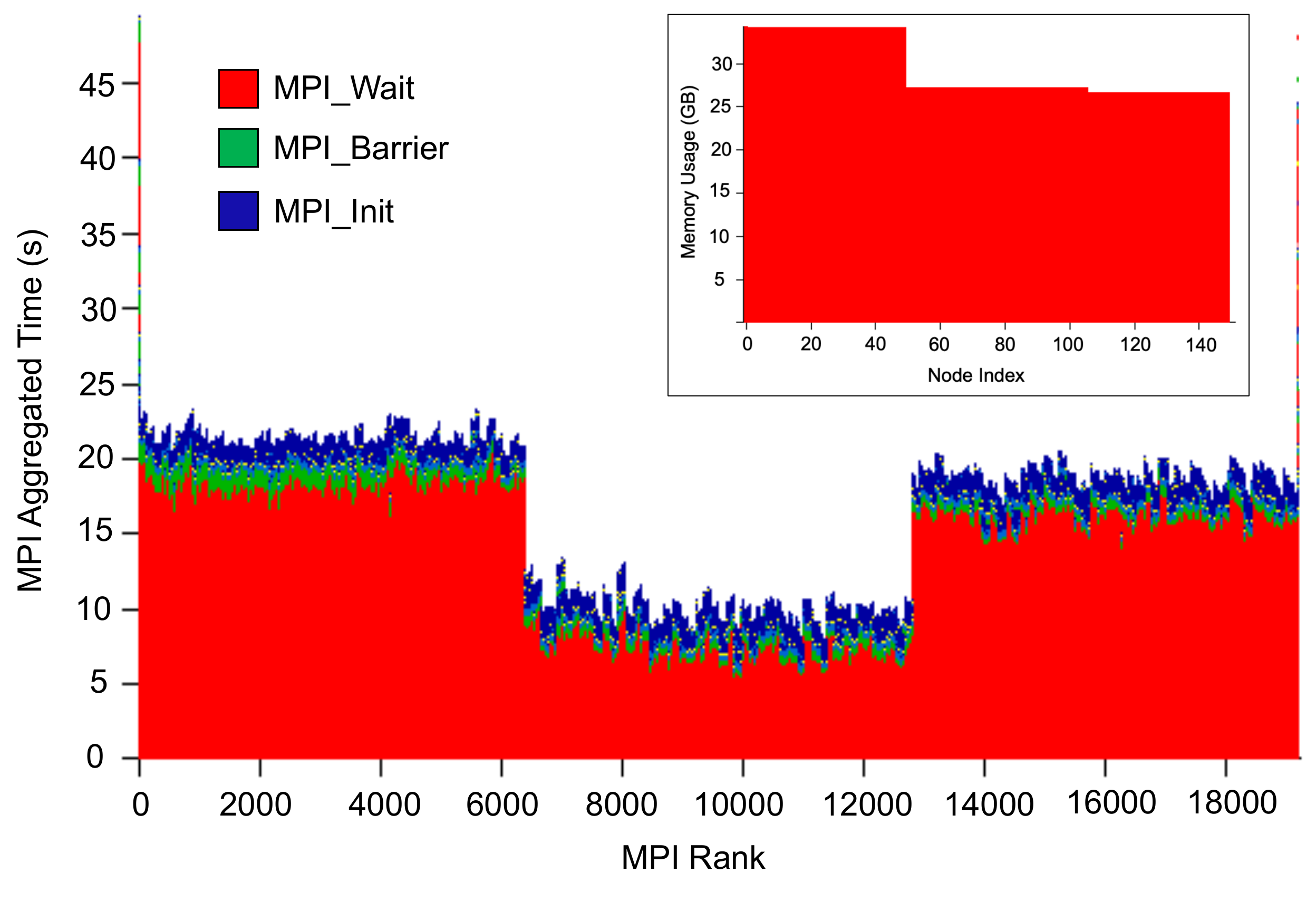}        \caption{MPI aggregated communication time for the BIT1 Sheath simulation on 19,200 cores. The insert shows the usage of memory per compute node (this plot shows an imbalance in memory usage).} \label{imbalance}
    \end{center}
\end{figure} 

We present these results in Fig.~\ref{imbalance}, where we have the cumulative time of MPI functions per rank. The MPI ranks at the domain boundaries (ranks \texttt{0} and \texttt{19199}) take more than double the time in MPI operations. The MPI time is dominated by the \texttt{MPI\_Wait} (in red color) synchronization function, showcasing a problem with load imbalance~\cite{peng2015cost}, as we have stated before. In addition, by inspecting Fig.~\ref{imbalance}, we can see that, excluding the boundary MPI processes, there are three large groups of MPI processes, which are \emph{self-synchronized}: two groups have an aggregated MPI call time of approximately 20 seconds, while one group has an aggregated MPI call time of approximately 10 seconds. The self-synchronization is due to the presence of idle waves~\cite{markidis2015idle}~\cite{peng2016idle} in a memory-bound MPI application with a simple communication pattern~\cite{afzal2021analytic}. In the Fig.~\ref{imbalance} insert, we show the amount of memory consumed per node. Part of the computing nodes has a more considerable use of the nodes, approximately 23\% more: the largest usage of memory per node is approximately 34~GB while the smallest is approximately 26~GB.  

\vspace{2mm} 

\highlight{\textbf{Observation V:} The \texttt{IPM} performance results confirm the workload imbalance issue with some processes spending more (MPI processes relative to domains at the boundaries) than double the time in MPI calls with respect to other MPI processes. At large number of MPI processes, BIT1 is subject to a self-synchronization process, degrading the overall parallel performance.}

\subsection{I/O Performance}
\begin{figure}[!h]
        \includegraphics[width=\textwidth]{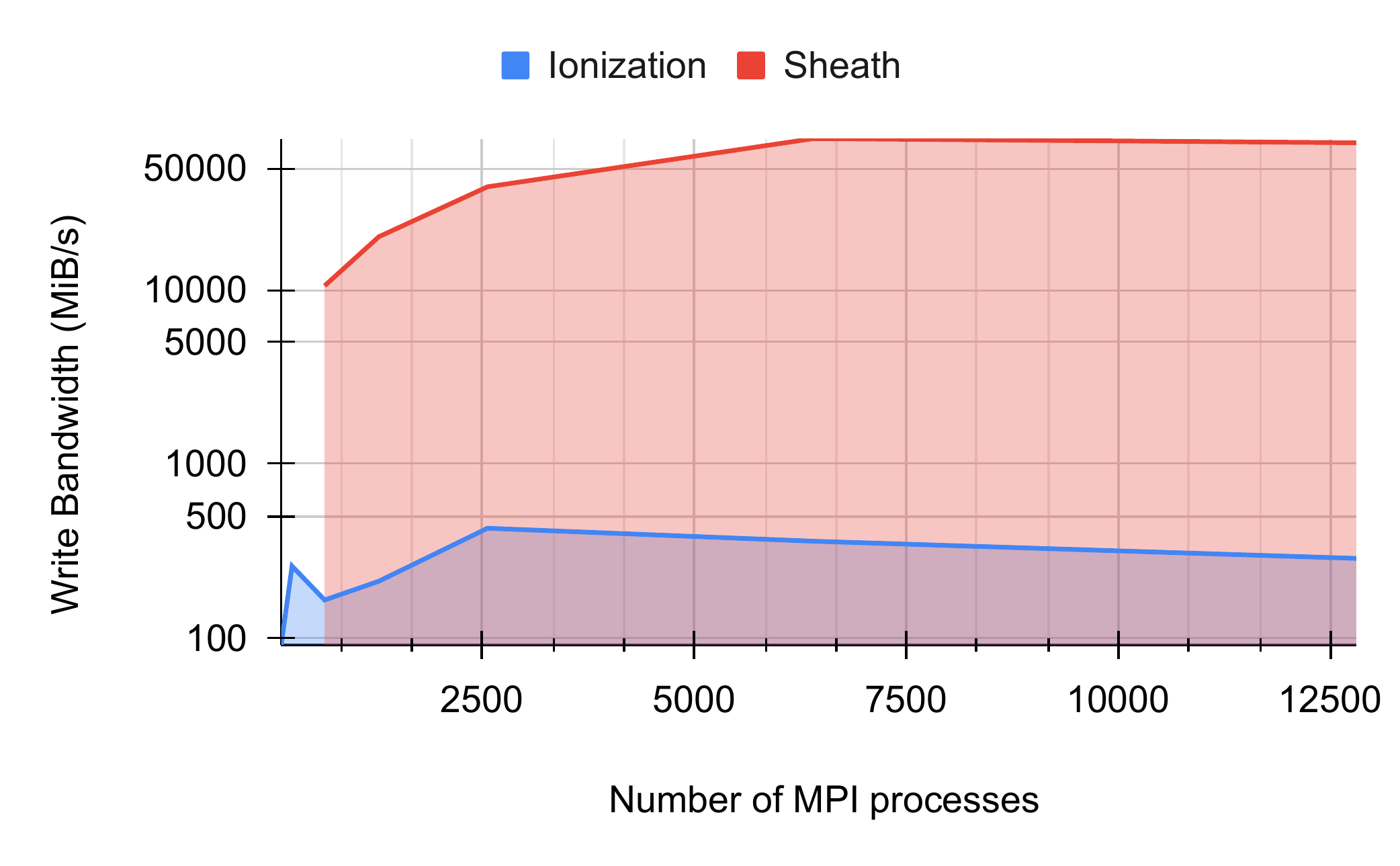}
        \caption{BIT1 I/O Write Bandwidth, measured in MiB/s} \label{darshan}
\end{figure}

Finally, we investigate the I/O performance when the diagnostics are activated. To better understand I/O performance, we use a simulation of 1,000 time steps. We have diagnostics output (with BIT1 I/O flags \texttt{slow} for plasma profiles and distribution functions, \texttt{slow1} for self-consistent atomic collision diagnostics, generating the required \texttt{.dat} files) every 100 cycles and checkpointing files (so-called \texttt{.dmp} files) every 333 cycles. The read operations are limited to read the simulation input files. We focus only on analysing the performance of the write operations. BIT1 performs serial I/O, e.g. each MPI process writes its own diagnostics and checkpointing files. Fig.~\ref{darshan} presents the write bandwidth results, measured with \texttt{Darshan} for the increasing number of MPI processes (up to 12,800 MPI processes) in the two test cases.
By studying the two plots, we observe that write bandwidth increases with the number of MPI processes, and then it saturates, for both the test cases. The write bandwidth saturates approximately at 300 MiB/s and 70 GiB/s for the Ionization and Sheath test cases, respectively. The peak I/O write bandwidth depends on the problem size. After the peak I/O is reached, the performance degrades as the metadata writing cost increases.
\vspace{3mm} 

\highlight{\textbf{Observation VI:} BIT1 performs serial I/O. The write bandwidth increases for a small number of MPI processes until the peak bandwidth is reached. For a large number of MPI processes, the I/O write bandwidth decreases as the cost associated with metadata write increases.}



\section{Discussion and Conclusion}
In this article, we presented the BIT1 code performance results and analysis, understanding the bottleneck and identifying optimization opportunities. 
We showed that BIT1 is a memory-bound code, and its performance is highly dependent on the problem size (that might fit or not to the different memory hierarchy capacities). Our scaling tests showed a superlinear speed-up that can be explained by better usage of LLC on a very large number of cores.

We have found that the main performance bottleneck is a highly memory-bound sorting function that depends on cache usage and available memory bandwidth. Future optimizations targeting the BIT1 performance must target the optimization of this function and consider the dependency on memory performance. At this end, using high-bandwidth memories~\cite{peng2016exploring} will likely increase considerably the BIT1 performance without any further code optimization.

The dependency of the code performance on the \texttt{arrj} sorting function (with many memory transfers and little to no computing) does not make BIT1 a good candidate to port to GPU systems. For porting BIT1 to GPU systems and achieving high performance, the BIT1 sorting code algorithm, or the data layout should be likely reformulated to reduce the memory traffic and increase the computational intensity to use GPUs effectively.

Another performance bottleneck is the imbalance that slows the simulation because of the synchronization of MPI point-to-point communication. We have found that MPI processes at the domain boundaries (dealing with the boundary conditions to the simulations) are considerably slower than other MPI processes. However, other processes must wait for the processes at the domain boundaries as they are bound to local synchronization of the message passing. Additionally, we have found that when utilizing a large number of cores, BIT experiences self-synchronization issues, leading to a degradation in overall parallel performance. Future MPI communication optimization strategies must address this issue and de-synchronize the MPI processes.

Finally, BIT1 would benefit from the usage of parallel I/O communication (such as utilizing MPI I/O or HDF5 parallel I/O libraries) to reduce the cost of metadata writing (as thousands of files are written at each I/O cycle) and enhance the I/O performance. Another possibility is to provide BIT1 with in-situ visualization and data analysis via ADIOS-2 and Kitware Catalyst libraries to reduce the overall cost of I/O and post-processing.
\space

\vspace{2mm} 
\noindent \small{\textbf{Acknowledgments.} Funded by the European Union. This work has received funding from the European High Performance Computing Joint Undertaking (JU) and Sweden, Finland, Germany, Greece, France, Slovenia, Spain, and Czech Republic under grant agreement No 101093261.} 



%
%
\bibliographystyle{splncs04}

%

\end{document}